\pdfoutput=1

\PassOptionsToPackage{x11names}{xcolor} % For review purposes only
\documentclass[pdflatex,iicol,sn-mathphys]{sn-jnl-fix}

\usepackage{fix-cm}
\usepackage{siunitx}
\usepackage[capitalise]{cleveref}
\usepackage{lineno} % For review purposes only

\newcommand*{\papertitle}{Performance of a convolutional autoencoder designed to remove electronic noise from p-type point contact germanium detector signals}

\newcommand*{\papertitleshort}{Performance of a convolutional autoencoder designed to remove electronic noise}

\newcommand*{\paperkeywords}{
    denoising,
    deep learning,
    neural networks,
    autoencoders,
    germanium detectors}

\hypersetup{
    pdfauthor={Mark R. Anderson},
    pdftitle={\papertitle},
    pdfkeywords={\paperkeywords}}

\newcommand{\of}[1]{\left(#1\right)}
\newcommand{\znubb}{$0\nu\beta\beta$}

\definecolor{idcolor}{HTML}{A6CE39}

\newcommand{\properwidth}{\columnwidth}

\ifx\orcid\undefined
    \newcommand{\orcid}[1]{}
    \newcommand{\linebreaks}{\newline\newline\newline}
\else
    \newcommand{\linebreaks}{}
    \linenumbers
\fi

\jyear{2022}

\begin{document}

%%============================================%%
%% Title page, including authors and abstract %%
%%============================================%%

\title[\papertitleshort]{\papertitle}

\author*[1]{\fnm{Mark} R. \sur{Anderson}~\orcid{0000-0003-4321-0094}}\email{anderson.mark@queensu.ca}
\author[1]{\fnm{Vasundhara} \sur{Basu}~\orcid{0000-0002-8388-8484}}
\author[1]{\fnm{Ryan} D. \sur{Martin}~\orcid{0000-0001-8648-1658}}
\author[1]{\fnm{Charlotte} Z. \sur{Reed}~\orcid{0000-0003-3406-7353}}
\author[1]{\fnm{Noah} J. \sur{Rowe}~\orcid{0000-0003-2407-4998}}
\author[1,2,3]{\fnm{Mehdi} \sur{Shafiee}~\orcid{0000-0003-3293-7191}}
\author[1]{\fnm{Tianai} \sur{Ye}~\orcid{0000-0002-5706-1459}}

\affil[1]{\orgdiv{Department of Physics, Engineering Physics \& Astronomy},
    \orgname{Queen’s University},
    \orgaddress{\city{Kingston},
                \state{ON},
                \country{Canada}}}

\affil[2]{\emph{Present Address:}
    \orgdiv{Department of Electrical and Computer Engineering},
    \orgname{Nazarbayev University},
    \orgaddress{\city{Nur-Sultan}, \country{Kazakhstan}}}

\affil[3]{\emph{Present Address:}
    \orgdiv{Energetic Cosmos Laboratory},
    \orgname{Nazarbayev University},
    \orgaddress{\city{Nur-Sultan}, \country{Kazakhstan}}}

\abstract{We present a convolutional autoencoder to denoise pulses from a p-type point contact high-purity germanium detector similar to those used in several rare event searches. While we focus on training procedures that rely on detailed detector physics simulations, we also present implementations requiring only noisy detector pulses to train the model. We validate our autoencoder on both simulated data and calibration data from an $^{241}$Am source, the latter of which is used to show that the denoised pulses are statistically compatible with data pulses. We demonstrate that our denoising method is able to preserve the underlying shapes of the pulses well, offering improvement over traditional denoising methods. We also show that the shaping time used to calculate energy with a trapezoidal filter can be significantly reduced while maintaining a comparable energy resolution. Under certain circumstances, our denoising method can improve the overall energy resolution. The methods we developed to remove electronic noise are straightforward to extend to other detector technologies. Furthermore, the latent representation from the encoder is also of use in quantifying shape-based characteristics of the signals. Our work has great potential to be used in particle physics experiments and beyond.
}


\maketitle

%%===========%%
%% Main text %%
%%===========%%

\section{Introduction}

High-purity germanium (HPGe) detectors are used in the search for rare events such as neutrinoless double-beta decay (\znubb) \cite{alvis2019search,agostini2020search,abgrall2021legend}, dark matter \cite{aalseth2013cogent,agostini2020first}, and other beyond Standard Model physics \cite{abgrall2016search,alvis2018first,alvis2019trinucleon}. HPGe detectors are a common choice due to their intrinsic purity, good energy resolutions, and ability to be fabricated from material enriched in the \znubb~candidate isotope $^{76}$Ge. Some of the strongest limits on the \znubb~half-life have been set using HPGe detectors and are in the range of $\SI{e25}{yr}$ - $\SI{e26}{yr}$ \cite{alvis2019search,agostini2020search}.

Due to the infrequent occurrence of signal events, extraordinary measures are taken to reduce backgrounds associated with particles that can deposit energy in the detectors, including locating the experiments in ultra-clean laboratories deep underground \cite{abgrall2014majorana}. Further analytical background reduction and discrimination techniques are still required as experiments probe larger regions of the parameter space associated with the processes that they are trying to detect.

An efficient denoising algorithm can help advance searches for rare event interactions. Noise reduction techniques can allow one to identify low-energy signal events that would otherwise be dominated by electronic noise. This is of direct relevance to experiments using germanium detectors to search for rare events at low energies, such as solar axions, violation of the Pauli Exclusion Principle, and electron decay \cite{abgrall2017new}. A reduction in noise would also allow for better background rejection techniques that are based on pulse shapes, such as the rejection of slow energy-degraded pulses in germanium detectors searching for signals at low energies \cite{aguayo2013characteristics,wiseman2020low}.

Denoising could also provide more accurate measurements of pulse amplitudes, leading to a better energy resolution. Many pulse height estimation algorithms \cite{jordanov1994digital,agostini2015improvement} use effective averaging windows with a given shaping time. While an overall reduction in the energy resolution is difficult to achieve compared to these highly efficient algorithms, denoising the pulses beforehand can reduce the shaping time required to obtain a comparable energy resolution. This can allow for shorter traces to be collected (more efficient data storage), a smaller sampling period to be used (more detailed pulses), and/or a higher data collection rate (lower energy thresholds).

In this paper, we demonstrate the effectiveness of deep learning to strongly reduce electronic noise in the charge pulses from a p-type point contact (PPC) HPGe detector. We conduct several studies on both real and simulated data to verify the performance of our deep learning-based model and compare it to traditional noise reduction techniques. We also show that an effective noise reduction model can be trained using \emph{only noisy detector data}. Since this method does not require underlying clean pulses for training, our approach can be implemented \emph{without} detailed modelling and simulations of the detector. Furthermore, our results are not limited to PPC HPGe detectors; this efficient and flexible denoiser which requires only detector data is widely applicable to the particle astrophysics community and beyond.

\Cref{sec:autoencoders_for_denoising} provides an overview of both traditional denoising methods and autoencoders. In \cref{sec:model}, we describe the design of our neural network architecture. In \cref{sec:experimental_setup}, we outline the experimental setup and datasets used for training and evaluation, while in \cref{sec:methodology} we describe our methodology. The results are presented in \cref{sec:results}, first on simulations to demonstrate both pulse shape preservation and improvements over traditional methods, and then on real detector data to further validate pulse shape preservation and to show the impact of denoising on the energy resolution. We conclude our results and emphasize the potential of our work for other experiments in \cref{sec:conclusion}.
\section{Autoencoders for denoising}
\label{sec:autoencoders_for_denoising}

\subsection{Traditional denoising methods}
\label{sec:traditional_denoising_methods}

We briefly discuss non deep learning-based approaches to denoise signals before introducing autoencoders. We focus on a subset of techniques that were evaluated for comparison with the autoencoder in \cref{sec:denoising_methods_comparison}. The first one is a moving average filter over $w$ samples. The selection of $w$ requires a trade-off between the level of noise reduction and the preservation of details such as edge sharpness in the pulse shape. An extension to the moving average is the exponential filter \cite{brown1956exponential}, where the smoothed output of a given sample is the weighted sum of the current sample and the previous prediction. The algorithm is recursive and is defined by the smoothing constant $\alpha \in (0,1)$.

We also consider the more advanced Savitzky-Golay filter \cite{savitzky1964smoothing}, which removes noise by fitting a degree-$p$ polynomial to $w$ adjacent samples centred about a given point of the signal and evaluating the fitted polynomial at this point. The Savitzky-Golay filter can be implemented as a weighted moving average with coefficients that depend on $w$ and $p$ \cite{savitzky1964smoothing}.

Another method we investigated applies thresholding rules to the wavelet decomposition of the noisy signal \cite{taswell2000wavelet}. Wavelet-based denoising requires a choice of mother wavelet and order as well as thresholds for the wavelet coefficients \cite{shafiee2016}. \Cref{tab:wavelet} lists the set of mother wavelets that were explored. Additionally, we tested two methods of thresholding: VisuShrink \cite{donoho1994ideal}, which applies a global threshold to the wavelet coefficients, and BayesShrink \cite{chang2000adaptive}, which determines thresholds at each subband of the wavelet by minimizing Bayesian squared error risk. For each method, we considered hard thresholding (keeping the coefficient if greater than the threshold) and soft thresholding (shrinking the coefficient toward zero by the threshold).

\begin{table*}[ht]
\centering
\caption{Wavelet functions used for denoising in this work, including the mother wavelet and order (where applicable).}
\label{tab:wavelet}
\begin{tabular}{@{}ll@{}}
\toprule
\textbf{Mother Wavelet} & \textbf{Order} \\
\midrule
Haar & N/A \\
Daubechies & 2 -- 38 \\
Coiflet & 1 -- 17 \\
Symlet & 2 -- 19 \\
Biorthogonal & 1.1, 1.3, 1.5, 2.2, 2.6, 2.8, 3.1, 3.3, 3.5, 3.7, 3.9, 4.4, 5.9, 6.8 \\
Reverse Biorthogonal & 1.1, 1.3, 1.5, 2.2, 2.6, 2.8, 3.1, 3.3, 3.5, 3.7, 3.9, 4.4, 5.9, 6.8 \\
Meyer & N/A (finite impulse response approximation) \\
\botrule
\end{tabular}
\end{table*}

The last technique examined in this work is the Kalman filter \cite{kalman1960linear}, which relies on knowledge of an underlying model of the data. Kalman filtering provides an estimate of the true value based on a model of the event itself and the measured samples. By knowing the accuracy of the event modeling technique as well as the measurement error, an interpolation can be made between the two.

\subsection{Autoencoders}

An autoencoder is an unsupervised machine learning algorithm used to encode data by learning from the data. The idea of the autoencoder was introduced in 1986 to learn an efficient coding for orthogonal inputs \cite{hinton1986learning}. The concept was further explored in the mid to late 1980s; for example, coupled hierarchical autoencoders were demonstrated to converge much faster on encoding a representation than a typical multi-hidden layer feedforward network at the time \cite{ballard1987modular}. Today, autoencoders are widely used for dimensionality reduction, anomaly detection \cite{sakurada2014anomaly}, and generative modelling \cite{kingma2014variational}. Autoencoders are also still frequently used for obtaining useful features or representations for other tasks. In \cite{holl2019deep}, for instance, an autoencoder was trained to learn a compressed representation of pulses from germanium detectors, and the encoded input was then used to train another network for pulse shape discrimination.

An autoencoder has two main components: the encoder is a function, $f$, that produces a latent representation of the input, while the decoder is another function, $g$, which takes the latent representation and generates a reconstruction of the original input. Using $x$ as the input, $y$ as the internal representation, and $z$ as the reconstructed output, the encoder transforms $x$ to $y$ by $f_{\theta}(x)=y$ and the decoder transforms $y$ to $z$ by $g_{\theta'}(y)=z$. The optimal parameters $\theta$ and $\theta'$ are selected by minimizing the reconstruction error $L$ between $x$ and $z$. Typically, the encoder and decoder are connected and trained as one model. In almost all references to autoencoders in the literature, both $f_{\theta}$ and $g_{\theta'}$ are neural networks, and further discussion herein will assume this. For a comprehensive review of neural networks and deep learning, including the core concepts of advanced architectures such as convolutional networks, see \cite{deeplearningnielsen} and \cite{Goodfellow-et-al-2016}.

\subsection{Denoising autoencoders}

\begin{figure}[ht]
	\centering
	
	\includegraphics[width=\properwidth]{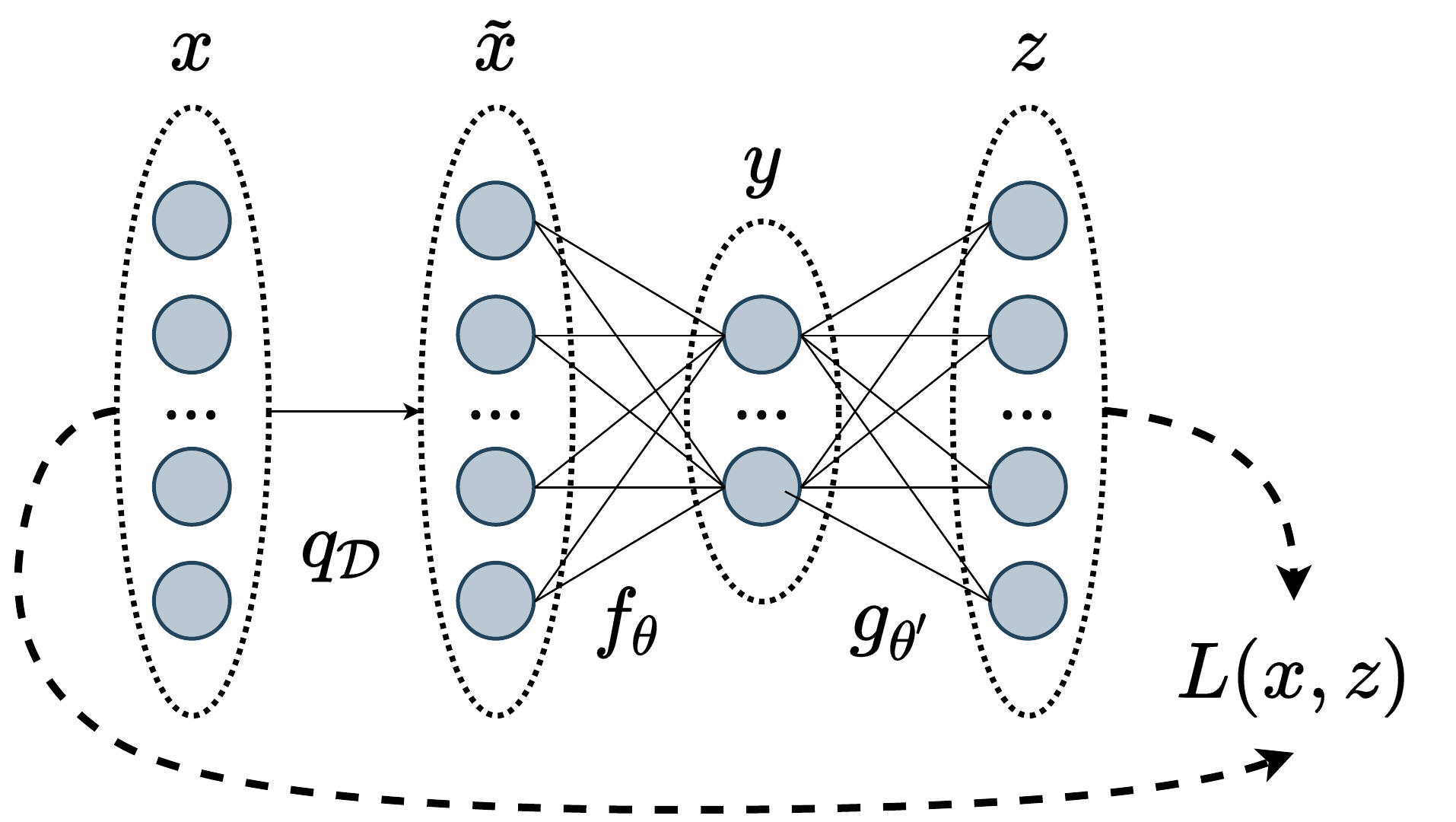}
	\caption{Basic concept of denoising autoencoders. An input $x$ is artificially corrupted by some noise process $q_{\mathcal{D}}$ to become $\tilde{x}$. The encoding portion of the autoencoder, $f_{\theta}$, produces a new representation $y$. The decoder portion of the network, $g_{\theta'}$, attempts to reconstruct the clean input. Its estimate is given by $z$, and a loss function, $L(x,z)$, quantifies the reconstruction. Notation is largely based off of figure from \cite{vincent2010stacked}.}
	\label{fig:denoise_autoencoder}
\end{figure}

Standard autoencoders can remove some of the input noise if the latent representation is highly compressed, as the network will be forced to extract only the most useful features. For example, in \cite{holl2019deep}, it was observed that the autoencoder removed some of the noise from the inputs, despite not being trained to do so. Denoising autoencoders take this a step further and make the objective to remove noise explicit. With denoising autoencoders, the input is instead an artificially corrupted version of the clean signal, which we denote by $\tilde{x}$. The encoder is thus a mapping $f_{\theta}(\tilde{x})$ of the noisy signal to the latent representation, and the decoder a mapping $g_{\theta'}(f_{\theta}(\tilde{x}))$ of the latent representation to a reconstruction of the clean signal. As the output is the same, the reconstruction error between $x$ and $z$, $L(x,z)$, is still minimized to train the network. The basic concept of the denoising autoencoder is illustrated in \cref{fig:denoise_autoencoder}.

Denoising autoencoders were originally proposed to extract robust features from the inputs \cite{vincent2010stacked,vincent2008extracting}. The primary goal was not to remove noise. Rather, denoising was used as a criterion to produce encoded representations which performed better on a variety of classification tasks.

Autoencoders specifically for denoising purposes have been successful in a variety of domains, although much of the existing research focuses on denoising images rather than one-dimensional signals. In \cite{jain2009natural}, convolutional networks are used to denoise images with both known and unknown Gaussian noise levels, achieving comparable or superior performance to state-of-the-art methods at the time. In \cite{xie2012image}, autoencoders are used for both denoising and removal of superimposed text on images, although a separate network is trained for each noise level.

More recent works take advantage of modern advances in deep learning to train convolutional networks to remove Gaussian noise, as well as other tasks \cite{zhang2017beyond}. Convolutional networks with noise estimation subnetworks have also been applied to more complex noise models \cite{guo2019toward}. Furthermore, denoising convolutional autoencoders have now been demonstrated to perform well on several practical applications in various fields, such as denoising medical images \cite{gondara2016medical}, ``repairing'' images of corrupted printed circuit boards \cite{khalilian2020pcb}, and speech enhancement \cite{zezario2018deep}.

\subsection{Noise2Noise}
\label{sec:background_noise2noise}

The Noise2Noise procedure is an approach used to train a denoising model without the need for clean target examples \cite{lehtinen2018noise2noise}, which, in our case, require detailed physics simulations. Although we outline the method here, we refer the reader to the original paper for details. The central idea is that the mean of the target distribution minimizes the sum of squared errors, or $L_2$ loss, between the target and the prediction. In the simple case of point estimation, the $L_2$ loss for a set of measurements $\{x_i\}$ and prediction $z$ is given by

\begin{equation}
    L_2 = \sum_i \|z - x_i\|_2^2.
\end{equation}

The best estimate $z$ that minimizes the $L_2$ loss is simply the mean, $z = \sum_i x_i$. One can observe that $z$ is unchanged as long as the mean of the measurements is unchanged. Given infinite data, the addition of zero-mean noise would produce the same estimate $z$. The same logic can be applied to the denoising task of minimizing the $L_2$ loss between the clean signals, $x_i$, and the output of the autoencoder, $z_i = g_{\theta'}(f_{\theta}(\tilde{x_i}))$:

\begin{equation}
    L_2 = \sum_i \|g_{\theta'}(f_{\theta}(\tilde{x_i})) - x_i\|_2^2.
    \label{eqn:l2_clean}
\end{equation}

If instead the target $x_i$ is replaced with corrupted versions of itself, $\hat{x}_i$, with samples drawn from a distribution or distributions with a mean of $x_i$, then, the optimal estimate $z_i$ remains unchanged. Put another way, the corruption of $x_i$ with zero-mean noise will produce the same estimate, given enough data. The loss function becomes

\begin{equation}
    L_2 = \sum_i \|g_{\theta'}(f_{\theta}(\tilde{x_i})) - \hat{x}_i\|_2^2,
    \label{eqn:l2_noise}
\end{equation}

where $\hat{x}_i$ is a different noisy realization of the same underlying signal. By minimizing \cref{eqn:l2_noise}, the autoencoder should learn to predict the mean of the distribution of noisy pulses. Intuitively, the task of mapping one version of a noisy pulse to another is impossible if the model is not over-parameterized, and the best it can do is predict the mean. With finite data, the above results are only approximately true. However, we show in this paper that the Noise2Noise approach, with some additions, can produce excellent results in practice.

\section{Model}
\label{sec:model}

\subsection{Inputs and outputs}

Signals are collected from a PPC HPGe detector described in \cref{sec:experiment_detector}. Each signal is a pulse that consists of $M$ voltages sampled at $\SI{8}{ns}$ intervals. $M$ can technically be any number, but is typically either 4096 or 8192 samples in our work. Thus, the inputs and outputs of the models are $M$-long vectors of voltage samples.

\begin{figure}[ht]
	\centering
	
	\includegraphics[width=\properwidth]{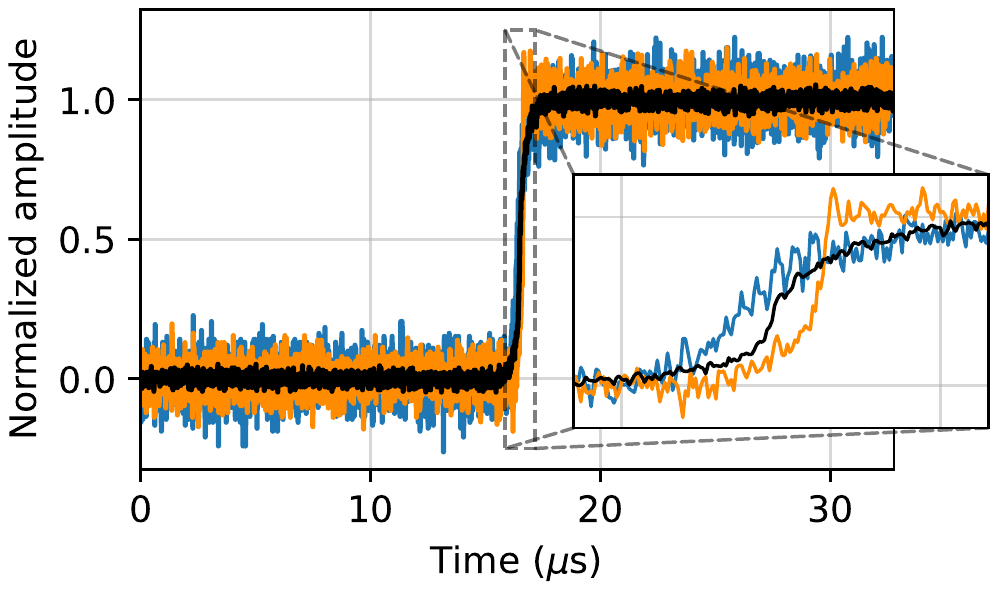}
	\caption{Three example pulses from a germanium detector, each with a substantially different rise time. All pulses are preprocessed such that they have a baseline of zero and an amplitude of one. The black pulse has apparently less noise because it was originally a higher amplitude pulse, so that the electronic noise is smaller relative to its amplitude.}
	\label{fig:data_example}
\end{figure}

All data pulses are preprocessed to have unit amplitude using a trapezoidal filter \cite{jordanov1994digital}, as described in \cref{sec:methodology_preprocessing}. An example of three preprocessed detector pulses with different rise times -- indicating different event positions in the detector -- and different signal-to-noise ratios is shown in \cref{fig:data_example}. The full length pulses are plotted and the rise region between approximately $\SI{16}{\mu s}$ and $\SI{17}{\mu s}$ is highlighted in the inset.

\subsection{Network architecture}

\subsubsection{Motivation}

Our architecture is fully convolutional and thus benefits from superior computational and memory efficiency over fully connected networks \cite{Goodfellow-et-al-2016}. As well, the convolutional nature of our model allows for a variable length input pulse, regardless of the size of the pulses it was trained on.

The parameter sharing aspect of convolutional networks is particularly important to our work as it forces the network to learn to remove noise in a consistent manner across the entire pulse, emphasizing feature locality. It also ensures that the network is equivariant to shifts on the time axis and thus independent of the relative position of the pulse in the trigger window. Finally, weight sharing greatly reduces the number of parameters to train, effectively acting as a regularizer to prevent overfitting.

\subsubsection{Design}

For a general 1-dimensional convolution, the size of the output $O$ is given by

\begin{equation}
	O = \frac{W - K + 2P}{S} + 1,
	\label{eqn:convoutsize}
\end{equation}

where $W$ is the input size, $K$ is the filter or window size, $S$ is the stride length, and $P$ is the padding. For a stride length or window size greater than one, the size of the output of the convolution may be different from the input, which has consequences on the selection of layer parameters. While the output size can be forced by padding the input, it is not apparent how to pad the pulses. Prepending or appending a constant value does not account for misalignment due to the imperfect normalization. Padding also ignores the noise.

An alternative approach is to use transposed convolution layers in the decoder to ``undo'' the size changes caused by non-padded convolutions. The transposed convolution operation uses the transpose of the convolution matrix -- a sparse matrix containing elements of the filter for computing the convolution using matrix multiplication -- to switch the forward pass with the backward pass \cite{dumoulin2016guide}. This effectively acts as a form of upsampling.

The architecture of our autoencoder is described in \cref{tab:architecture}, with each layer consisting of its stride, window, and output size. The first element of the output is the temporal length after the operation is applied, while the second element is the number of filters. The batch size is not included in the output shape. For illustration, the table uses a fixed input length of 4096 samples. A different input shape will only change the temporal dimension of the output shape as the number of filters does not depend on the input.

\begin{table}[ht]
\centering
\caption{Summary of the convolutional autoencoder architecture used in this paper for a fixed pulse size of 4096. Included in this table is the type, strides length $S$, window size $K$, and output shape $O$ of each layer.}
\label{tab:architecture}
\begin{tabular}{@{}llll@{}}
\toprule
\textbf{Layer} & \textbf{Stride} & \textbf{Window} & \textbf{Output} \\
\midrule
Input &  &  & 4096, 1 \\
Convolution & 1 & 1 & 4096, 8 \\
Convolution & 1 & 9 & 4088, 16 \\
Average Pooling & 2 & 2 & 2044, 16 \\
Convolution & 1 & 17 & 2028, 32 \\
Average Pooling & 2 & 2 & 1014, 32 \\
Convolution & 1 & 33 & 982, 64 \\
Average Pooling & 2 & 2 & 491, 64 \\
Convolution & 1 & 33 & 459, 32 \\
\midrule
Trans. Convolution & 1 & 33 & 491, 32 \\
Upsampling & 2 & 2 & 982, 64 \\
Trans. Convolution & 1 & 33 & 1014, 64 \\
Upsampling & 2 & 2 & 2028, 64 \\
Trans. Convolution & 1 & 17 & 2044, 32 \\
Upsampling & 2 & 2 & 4088, 32 \\
Trans. Convolution & 1 & 9 & 4096, 16 \\
Convolution (output) & 1 & 1 & 4096, 1 \\
\midrule
\multicolumn{4}{c}{\textbf{Total number of parameters: 286,145}} \\
\botrule
\end{tabular}
\end{table}

The network begins with an eight-filter convolution operation with a stride length and window size of one. This does not change the temporal dimension of the input, but rather acts as a sort of ``preprocessing'' layer that increases the overall dimensionality of the signal. The decoder also ends with a convolution layer where the stride length, window size, and number of filters are set to unity to recover the original shape of the pulse.

Both the encoder and decoder have three blocks of layers. A ``block'' consists of a convolution followed by a two-fold average pooling operation or a two-fold upsampling operation followed by a transpose convolution for the encoder and decoder respectively. The encoder compresses the original input by a factor of approximately eight, due to both the downsampling operation and \cref{eqn:convoutsize}. Every convolution and transpose convolution layer \emph{except} for the last is followed by a rectified linear unit (ReLU) activation function \cite{glorot2011deep}. No activation is applied to the final convolution layer as pulses can have values outside of the range (0, 1) or (-1, 1), and bounding the output was found not to be necessary.

The window sizes are chosen to \emph{increase} for each layer in the encoder and \emph{decrease} for each layer in the decoder. The window sizes were all chosen to be smaller or approximately the same size as a typical rise time. However, they are also subject to some restrictions. This can be seen from \cref{eqn:convoutsize}, which simplifies to

\begin{equation}
	O = W - K + 1
	\label{eqn:convoutsizesimp}
\end{equation}

for a stride length of one and no padding. If $W$ is even (odd), $K$ must be odd (even) in order for $O$ to always be even. $O$ must be even to ensure that the size of the input to the subsequent downsampling layer is divisible by two. The window size in each convolution layer, $K$, is selected such that this condition is true for each layer in the network starting with an input of 4096 samples. Conversely, pulses of variable length are subject to this requirement for the chosen window sizes.

\section{Experimental setup, data, and simulations}
\label{sec:experimental_setup}

\subsection{Detector}
\label{sec:experiment_detector}

Detector pulses were collected from a \SI{1}{kg} PPC HPGe detector manufactured by ORTEC/AMTEK \cite{ortec2022}. The detector is a cylinder with a radius of approximately \SI{3}{cm} and \SI{5}{cm} in height. The detector was operated in a PopTop cryostat, and preamplifier signals were recorded using a 16-bit \SI{125}{MHz} SIS3316 digitizer fabricated by Struck Innovative Systems \cite{struck2022}. The detector depletion voltage is \SI{2750}{V} and was operated at a bias of \SI{3700}{V}, as recommended by the manufacturer. Data were stored and preprocessed with a custom C++ analysis suite that is based on CERN's ROOT software \cite{brun1997root}.

\subsection{Real data}
\label{sec:experiment_detector_data}

\subsubsection{241-Americium source}

Data were collected from a $\SI{10}{\mu Ci}$ $^{241}$Am encapsulated source, which produces \SI{60}{keV} gamma rays. At this energy, electronic noise is a significant component of the pulse. By using a low energy collimated gamma source, the location of the collimator can be used to infer the location of energy depositions since Compton scattering is unlikely and the gamma rays do not typically penetrate more than \SI{1}{mm} in the detector. Furthermore, the interactions of these gamma rays in the detector are almost entirely single-site. Each trace is 8192 samples in length. Data were collected in December 2021.

\subsubsection{60-Cobalt source}

Data were also collected from a $^{60}$Co source that produces gamma rays with energies of \SI{1173}{keV} and \SI{1332}{keV}. In practice, data were collected over energies ranging from approximately $\SI{500}{keV}$ to $\SI{3}{MeV}$, including events from room backgrounds and the $^{60}$Co source. These data include many multi-site events from gamma rays Compton scattering in the detector. Each trace is 4096 samples in length. Data were collected in June 2020 and October 2020.

\subsubsection{Noise}

A large number of noise traces were collected from the detector in order to train the models with realistic electronic noise. These were obtained by randomly triggering the SIS3316 digitizer to read out signals from the PPC detector. Each trace is 4096 samples in length. Noise data were collected at three different times over a period of two and a half years: July 2019, January 2021, and December 2021. At each time point, the detector was under different operating conditions, resulting in a diverse noise set. Traces collected in December 2021 are 8192 samples in length, while the remaining sets are 4096 samples in length.

Signals in the noise data were filtered to remove any events that occasionally occur in the same trigger window. This was done by rejecting outliers based on the baseline and root mean square (RMS) of the baseline, calculated over the first 1000 samples of each trace. Similarly, pulses with outliers in the minimal or maximal values from a trapezoidal filter with a gap window of $\SI{1.8}{\mu s}$ and a rise window of $\SI{6}{\mu s}$ were removed from this dataset.

\subsection{Simulated data}
\label{sec:experiment_simulated_data}

\subsubsection{Library pulses}

A detailed physics-based simulation was used to create a set of $\num{1724}$ ``library'' pulses on a $\SI{1}{mm}\times\SI{1}{mm}$ grid in radius and height to represent pulses uniformly in the azimuthally symmetric PPC detector \cite{vasuthesis2020}. This position-dependent basis set was created using the \texttt{siggen} simulation software which models the propagation of charges in germanium detectors \cite{siggen}. The library pulses can be used to infer realistic clean signals underlying actual events in the detector using a $\chi^2$ minimization between a normalized data pulse and every library pulse in the basis set \cite{vasuthesis2020},

\begin{equation}
	\chi^2 = \sum_{j}^{M}\frac{\of{z_{k,j}-x_{i,j}}^2}{\sigma_i^2},
	\label{eqn:chi-square}
\end{equation}

where the $x_{i,j}$ is the $j^{\text{th}}$ sample of the $i^{\text{th}}$ data pulse, $z_{k,j}$ is the $j^{\text{th}}$ sample of the $k^{\text{th}}$ library pulse in the basis set, and the quantity is evaluated over $M$ samples. $\sigma_i$ represents the noise level in the data pulse, which we take as the RMS of the noise in the pre-trigger baseline.

Multiple sets of library pulses were generated with different preamplifier time constants to further expand this dataset. For a given position, this affects the rise time and the curvature of the pulse. Using preamplifier time constants of $\SI{0}{ns} - \SI{80}{ns}$ in increments of $\SI{10}{ns}$, the library pulse dataset was expanded to $\num{15516}$ unique traces.

\subsubsection{Piecewise linear smoothed pulses}

A set of pulses that look similar to the library pulses was generated by using piecewise mathematical functions. These pulses mimic the shape of the library pulses without requiring any complex physics simulation, and do not depend on the details of a specific detector. Each pulse was composed of two linear pieces connected at a varying fixed point to mimic the slow and fast portions of the rise. All pulses have an amplitude of unity and a rise time between $\num{25}$ samples and $\num{125}$ samples. The pulses were smoothed using a moving average with a window size of 5\% of the rise time. A total of $\num{20070}$ unique traces were created using this procedure. We refer to these signals as piecewise linear smoothed (PLS) pulses.

\begin{figure}[ht]
	\centering
	
	\includegraphics[width=\properwidth]{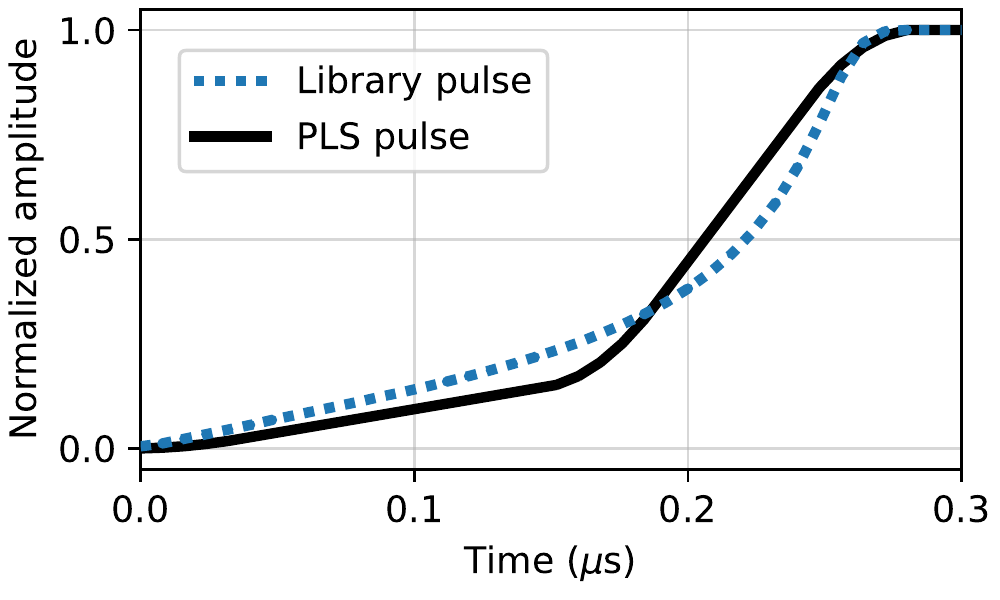}
	\caption{An example of a simulated PLS pulse (solid line) and library pulse (dotted line), each with a rise time of $\SI{280}{ns}$.}
	\label{fig:fake_example}
\end{figure}

\Cref{fig:fake_example} shows an example library pulse and PLS pulse, each with a short rise time of $\SI{280}{ns}$. The PLS pulses tend to be noticeably sharper than the library pulses and only roughly approximate their general shape.

\section{Methodology}
\label{sec:methodology}

\subsection{Datasets}

The library and PLS simulated pulses described in \cref{sec:experiment_simulated_data} were used as distinct datasets to train and evaluate the performance of the model. Pulses in these sets were padded to have 4096 samples, which was found to be sufficient for the network to denoise the flat regions of the pulses well without distorting the important components of the pulses.

Data from the $^{241}$Am source were used to validate the denoising algorithm by comparing denoised pulses to their known shapes from simulations. Since data from the $^{60}$Co source include a variety of high-energy, multi-site events, they provided an ideal set of diverse pulses for training the autoencoder with the Noise2Noise method.

\subsection{Data preprocessing}
\label{sec:methodology_preprocessing}

The baseline, defined as the mean over the first 1000 samples of a pulse, was calculated and subtracted from each signal. The pulse amplitudes were calculated by applying a trapezoidal filter to each trace \cite{jordanov1994digital}, and the baseline-removed pulses were then scaled by their amplitudes. Additionally, a pole-zero correction was applied to each pulse to remove the main component of the exponential decay from the resistive feedback preamplifier that was used to read out the detector. We refer to this entire preprocessing procedure as ``amplitude normalization'', noting that it includes the pole zero correction for data pulses (the simulated pulses do not have an exponential decay). Because of the noise, the amplitude normalized pulses roughly, but not exactly, range in height from zero to one.

We also explored standardizing each pulse to have a mean of zero and standard deviation of 0.5 after the pole-zero correction. A value of 0.5 was used to ensure that horizontally centred pulses have an amplitude of roughly one. We refer to this method of preprocessing as ``standardization'', noting again that it includes the pole zero correction for data pulses. However, models trained with standardized pulses were found to perform similarly or even slightly worse in most circumstances. Additionally, these models were found to depend heavily on the absolute position in the pulse where the rise region begins. Due to the lack of robustness to horizontal shifts, all models described in the results section can be assumed to have been trained with amplitude normalized pulses unless mentioned otherwise.

\subsection{Data augmentation}
\label{sec:data_augmentation}

\subsubsection{Multi-site event generation}

The set of library pulses form a basis of position-dependent signals in the detector and thus contain only single-site events. Similarly, the PLS pulses are simple mathematical functions constrained by two points and are analogous to single-site events in shape. To augment the training data, artificial multi-site events are created by adding randomized combinations of simulated pulses together, without mixing between the library and PLS sets.

Events are generated with up to five sites. This upper bound is chosen because for a Poisson process with an expected rate of two, the probability of a number greater than five is less than 2\%.  Furthermore, an equal number of events are generated for each number of sites, rather than being based on a physical distribution for the number of Compton scatters. The number of multi-site events is thus four times larger than the number of single-site events. We found this to be optimal for ensuring that the network does not smooth over multi-site events while also preserving the shape of single-site events.

In order to generate an $n$-site event, $n$ pulses are drawn from a set of simulated pulses. A random amplitude is drawn from a uniform distribution, $\sim U(0, 1)$, for each pulse. Each pulse is also horizontally shifted by a value drawn from a discrete uniform distribution, $\sim U(0, 100)$, to account for the possible drift times from different points in the detector. The scaled and shifted pulses are then added together to create an artificial multi-site event, and the result is rescaled to have an amplitude of one.

\subsubsection{Shifting and scaling}

The process of amplitude normalization on detector signals is imperfect due to noise. Simulated pulses, in contrast, are noiseless and have perfect normalization. Preliminary results showed that the network would overfit the beginning and the end of the predicted pulses if the clean pulses started at exactly zero and ended at exactly one. To combat this issue, random shifting and scaling is applied to all simulated pulses during training. This ensures that the network sees a variety of imperfect normalizations and is able to better generalize to real detector data while avoiding overfitting. This type of artificial data augmentation can also help improve the overall performance of the network by effectively providing more data than is available. In order of application, the shifting and scaling procedure consists of:

\begin{enumerate}
    \item \textbf{Amplitude scaling}: Each pulse is rescaled to have an amplitude drawn from the uniform distribution $\sim U(0.9, 1.1)$.
    \item \textbf{Vertical shifts}: A random vertical shift, drawn from a uniform distribution $\sim U(-0.1, 0.1)$, is applied to each pulse.
    \item \textbf{Horizontal shifts}: For each pulse of length 4096 samples, a number is drawn from a discrete uniform distribution $\sim U(1000, 3000)$, and the pulse is shifted such that the rise begins at this randomly chosen sample.
\end{enumerate}

While vertical scaling and shifting of $\pm10\%$ of the pulse amplitude is proportionally much larger than observed with typical data pulses, we use such a wide range of random variations to further improve generalization of the autoencoder, particularly to high noise pulses and outliers such as pile-up events. As well, horizontal shifts only have a major effect near the edges of an event due to the receptive field of the convolution layers. Thus, the unrealistic range of values used in this procedure is primarily for data augmentation purposes. We also verified that shifting and scaling parameters can be selected differently to favour performance on outliers.

\subsubsection{Noise addition}

Noise collected from the detector is added to the clean pulses after the previous data augmentation steps are applied. The detector noise dataset is sufficiently large such that each clean pulse has a unique, randomly chosen noise trace. Thus, noise seen in training is not seen in the validation and test phases, and noise pulses are not shared between datasets. All detector noise pulses are standardized to have a mean of zero and a standard deviation of $\sigma$.

In order to understand the effect of real detector noise, zero-mean normally distributed noise with no covariance over time, $\sim \mathcal{N}(0, \sigma)$, is also used, separately, to create independent datasets for comparison. For both noise types, $\sigma$ is drawn randomly for each trace from a uniform distribution, $\sim U(0, 0.2)$, to simulate the effect of varying the signal-to-noise ratio.

\subsection{Training procedures}

Two training procedures are explored in this paper: a regular training procedure which requires a noisy input and clean output, and a procedure which does not require the clean version of the noisy pulse. The latter procedure uses and extends the Noise2Noise approach developed in \cite{lehtinen2018noise2noise} and summarized in \cref{sec:background_noise2noise}.

The networks are implemented using the Keras API from TensorFlow \cite{chollet2015keras, tensorflow2015-whitepaper}. Training, validation, and testing were done using NVIDIA GeForce GTX Titan X and GeForce RTX 3090 GPUs.

\subsubsection{Regular training}
\label{sec:methodology_train_proc_reg}

The regular training procedure consists of applying the augmentation recipe above to a set of simulated data $d$ times, resulting in $N = d \times 5N_0$ pulses from an initial set of $N_0$ clean, single-site pulses. This is done separately with library and PLS pulses to understand whether detailed physics simulations are required for training the algorithm. The augmentation procedure is also applied twice per simulated set using Gaussian noise and real noise from the detector, for a total of four distinct datasets. We set $d=45$ for the library pulses and $d=35$ for the PLS pulses to ensure that the sizes of each training set are comparable. A copy of the pulses after the shifts and scales, but before the noise addition, is created and used for the clean targets.

The network is then trained to map the noisy pulses to the corresponding clean pulses. We use the Adam algorithm -- a stochastic gradient-based adaptive optimization procedure \cite{kingma2014adam} -- with a learning rate of $\num{3e-4}$ to minimize the mean squared error between the true clean pulse and the denoised pulse. We use a batch size of 128, which is large enough to ensure that the network trains reasonably quickly while not being so large as to affect the convergence of the network. Training is conducted over 100 epochs.

For each of the augmented simulated datasets, 10\% of the data are withheld for testing. Of the remaining 90\%, 10\% of that are withheld for validation and the remainder are used for training. At every epoch of training, the network is run on the validation set and the model at the epoch with the lowest loss is saved. The validation set is also used to select the best hyperparameters, including the batch size, learning rate, and architecture details. The best model -- as determined by the validation set -- is then run on the reserved test set. All results shown in this paper are calculated only on the test set.

\subsubsection{Noise2Noise training}

Unlike the regular training procedure, the Noise2Noise approach can be applied to \emph{both} simulated and raw detector data. The procedure follows the regular training procedure with the amendment of adding noise to the target pulse as well. For real detector data, where no clean underlying pulse exists, we add noise to the already noisy signals. Of course, the optimal solution to the minimization problem is then the underlying noisy pulse rather than the true clean pulse. To alleviate this issue, we add a simple penalty on the total variation of the denoised signal to the loss function $L$,

\begin{equation}
    L = L_0 + \frac{\lambda}{N}\sum_{i}^{N}\sum_{j}^{M-1}\lvert z_{i,j+1} - z_{i,j}\rvert,
    \label{eqn:smoothness_penalty_loss}
\end{equation}

where $L_0$ is the original loss function (which in our case is the $L_2$ loss), $z_{i,j}$ is the $j^{\text{th}}$ sample of the $i^{\text{th}}$ denoised output pulse, $M$ is the number of samples in the pulse, $N$ is the size of the training set, and $\lambda$ is a scaling factor. Minimization of the total variation, with some criterion of similarity to the original signal, was introduced as a method to denoise signals in 1992 \cite{rudin1992nonlinear}. Total variation is particularly useful in retaining the important components of a signal, including sharp edges and discontinuities, while avoiding the overcompensated smoothing present in many traditional denoising techniques \cite{strong2003edge}. Since the true pulses are monotonically increasing functions, the regularizer in \cref{eqn:smoothness_penalty_loss} will evaluate to the amplitude in the case of perfect denoising. If the denoised pulse still contains components of the noise and is very jagged, the penalty will become large.

The augmentation procedure for the simulated datasets is the same, except that a different noise pulse is also added to the target pulse. In the augmentation procedure for detector data, the artificial multi-site event generation step is skipped as the $^{60}$Co dataset used in training already contains multi-site events. As well, pulses are not horizontally shifted because the network is already equivariant to such shifts and because it is difficult to handle the bounds. However, the remainder of the procedure is still applied because the $^{60}$Co dataset consists of primarily high energy events, meaning that the shifts, scales, and noise are unrepresentative of lower energy pulses and outliers. For the noise addition, $\sigma$ is instead drawn from the uniform distribution $\sim U(0.025, 0.25)$. These numbers are chosen so that at least some noise is added to the data pulses, and to minimize correlated noise between the inputs and the targets.

The network is then trained to map one version of a noisy pulse to another. We again use the Adam algorithm to minimize the loss in \cref{eqn:smoothness_penalty_loss}. We use the validation set of the augmented library pulses with detector noise to select the best model as the data pulses do not have a target pulse. We specifically choose the model which minimizes the mean squared error, not the total loss, as the total variation penalty tends to dominate the validation loss. A batch size of 128 is still used, although we set the learning rate to $\num{e-5}$ as the learning rate in the regular training procedure was found to be too large.

\section{Results}
\label{sec:results}

\subsection{Evaluation on simulated data}

\subsubsection{Comparison between data augmentation and training procedures}

We evaluate the performance of the denoising convolutional autoencoder trained with several different datasets for the two training procedures. For each model, we test its generalization on the four available simulated datasets. The results are shown in \cref{tab:results_sim}. The first three columns are related to the training method and show the procedure, the dataset, and the type of additional noise used in training, respectively. The remaining columns show the mean squared error evaluated by denoising the simulated test datasets, containing either Gaussian or real detector noise on single- and multi-site events generated using either the library or the PLS pulses, for which the clean target is known. All pulses in the test sets were amplitude normalized and horizontally centred before denoising and comparison.

\begin{table*}[ht]
\centering
\caption{Summary of the mean squared error on the test sets for different models. The procedure, dataset, and noise type used in the training are given.}
\label{tab:results_sim}

\begin{tabular}{@{}lllllllll@{}}
\toprule

\multicolumn{3}{c}{\textbf{Training procedure and data}} & & \multicolumn{5}{c}{\textbf{Mean squared error} ($\times \num{e-5}$)} \\ \cmidrule{1-3} \cmidrule{5-9}
		
& & & & \multicolumn{2}{c}{\textbf{Gaussian noise}} & & \multicolumn{2}{c}{\textbf{Detector noise}} \\ \cmidrule{5-6} \cmidrule{8-9}
		
\textbf{Procedure} & \textbf{Data} & \textbf{Noise} & & \textbf{Library} & \textbf{PLS} & & \textbf{Library} & \textbf{PLS} \\

\midrule
Regular & Library & Detector & & 4.12 & 4.72 & & 3.76 & 4.21 \\
Regular & Library & Gaussian & & 3.40 & 3.82 & & 4.50 & 4.77 \\
Regular & PLS & Detector & & 5.10 & 4.48 & & 4.15 & 3.57 \\
Regular & PLS & Gaussian & & 3.93 & 3.36 & & 5.02 & 4.31 \\
\midrule
N2N ($\lambda = 0$) & Library & Detector & & 3.90 & 4.37 & & 3.86 & 4.20 \\
N2N ($\lambda = 0$) & Library & Gaussian & & 3.46 & 3.87 & & 4.57 & 4.82 \\
N2N ($\lambda = 0$) & PLS & Detector & & 5.11 & 4.48 & & 4.14 & 3.55 \\
N2N ($\lambda = 0$) & PLS & Gaussian & & 3.85 & 3.46 & & 4.97 & 4.43 \\
\midrule
N2N ($\lambda = 0$) & Detector & Detector & & 6.54 & 6.30 & & 7.78 & 7.40 \\
N2N ($\lambda = \num{e-2}$) & Detector & Detector & & 4.17 & 4.54 & & 5.04 & 5.26 \\
\botrule

\end{tabular}
\end{table*}

Each model trained on simulated data tends to perform best on the same class of pulses that it was trained on. Models trained with Gaussian noise tend to generalize slightly worse to detector noise than the other way around. As well, models trained using library and PLS simulated pulses have similar performance. However, models trained using library pulses tend to generalize better to PLS pulses than the other way around. This is evidenced by the difference between the performance on the test set of library and PLS pulses for a given noise type; the gap is larger for the model trained on PLS pulses.

The Noise2Noise models trained on simulated pulses perform nearly identically on the test set as the corresponding regular models in most cases. Differences in the mean squared error are typically on the order of $\num{e-7} - \num{e-6}$, which is expected due to statistical fluctuations in training. No total variation penalty was used as it was not found to improve the performance in the case of simulations. On simulated pulses with detector noise, neither Noise2Noise model trained on pulses from the $^{60}$Co source outperforms any of the models trained with simulated data, regardless of the procedure. However, the Noise2Noise model is still effective at denoising and requires only noisy detector data. A larger set of detector data, including from other high-energy sources which produce multi-site events, could improve the performance of this model.

Furthermore, we evaluated the performance of the Noise2Noise model with and without a total variation penalty and note that unlike the models trained on simulated data, the total variation penalty has a non-negligible impact on the denoising performance. We selected a penalty of $\lambda = \num{e-2}$ for \cref{tab:results_sim} as it was found to have the best mean squared error on the validation set, although using any penalty in the range $\num{e-4}$ -- $\num{e-1}$ produced similar results. The mean squared error is approximately 30\% lower by setting $\lambda$ in this range.

\begin{figure}[ht]
    \centering
    
    \includegraphics[width=\properwidth]{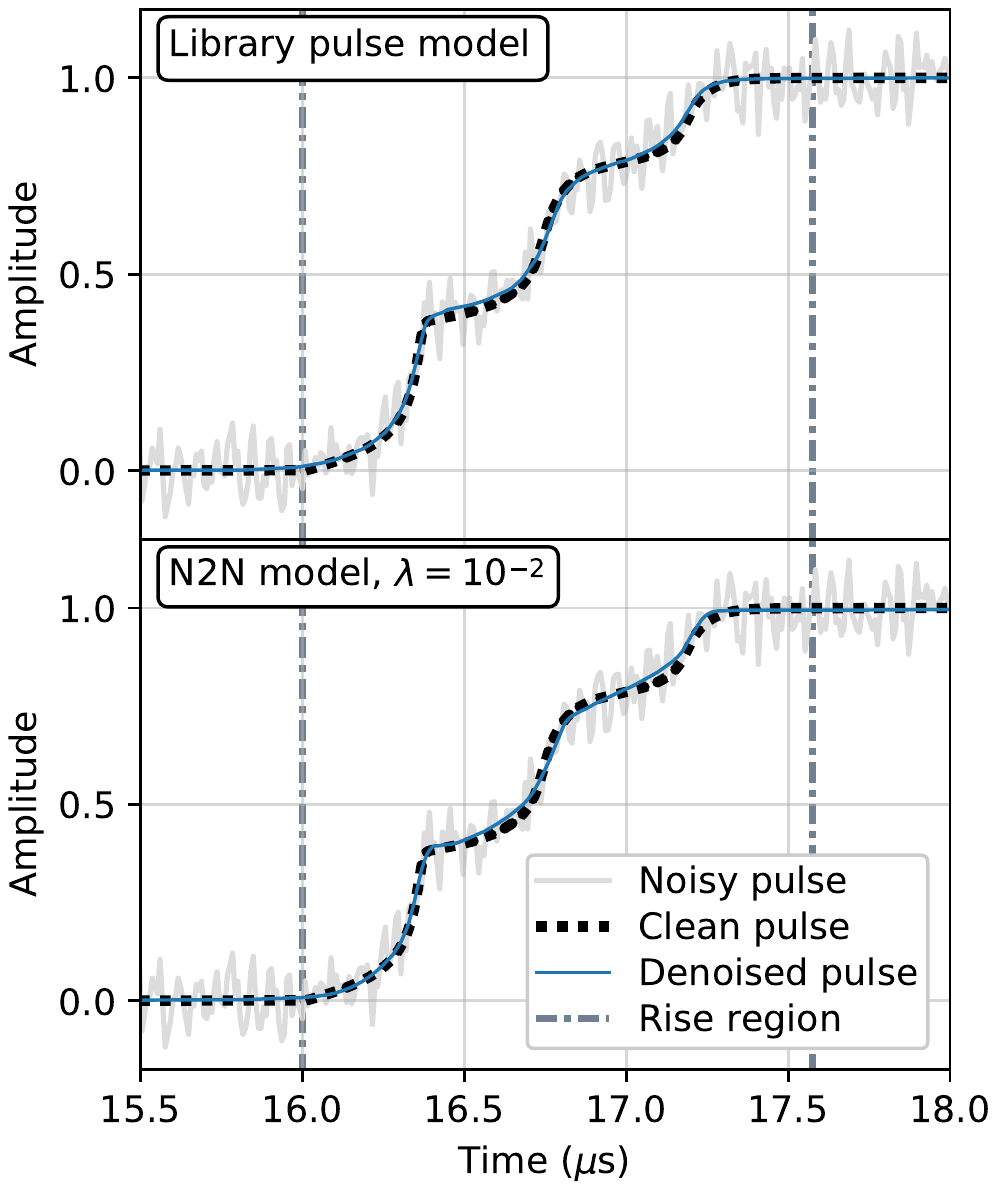}
    \caption{An example multi-site event from the library dataset. Included in each plot is the simulated pulse with artificial noise (solid light line), the clean underlying pulse (dotted line), and the corresponding denoised pulse (solid dark line) from the regular library pulse model (top) and Noise2Noise model (bottom).}
    \label{fig:lib_pulses_examples_denoised}
\end{figure}

\Cref{fig:lib_pulses_examples_denoised} shows an example library pulse multi-site event that has been denoised by two different versions of the autoencoder. The top panel shows the pulse denoised with the regular library pulse model while the bottom panel shows the denoised pulse using the Noise2Noise model trained with $^{60}$Co data with a total variation penalty.

Qualitatively, the regular library pulse model appears to fit the true underlying pulse the best, in line with the results in \cref{tab:results_sim}. However, the differences are subtle, and the Noise2Noise model appears to remove most of the noise without much additional distortion, and can be suitable for many physics applications that do not require highly detailed shapes to be preserved. Furthermore, although not shown here, visualization of the denoised pulses illustrates the impact of including a total variation penalty for the Noise2Noise method, as it removes much of the noise still present in the pulse denoised without it.

We emphasize that while all methods perform well, only the regular training procedure with library pulses required careful simulations of the detector.

\subsubsection{Comparison to traditional denoising methods}
\label{sec:denoising_methods_comparison}

The denoising performance of the autoencoder was compared to the traditional noise removal methods described in \cref{sec:traditional_denoising_methods}. The mean squared error between the clean target and denoised output was used as the metric and evaluated on the simulated library pulse test set. The performance was evaluated over the entire pulse as well as over two distinct sections of the pulse: the rise region and flat region. The rise region begins where the simulated pulse deviates from the flat baseline and ends where the pulse reaches its maximum amplitude. The flat region is defined as the area outside of the rise region. The example pulse in \cref{fig:lib_pulses_examples_denoised} includes an illustration of the region boundaries.

\begin{figure}[ht]
    \centering
    
    \includegraphics[width=\properwidth]{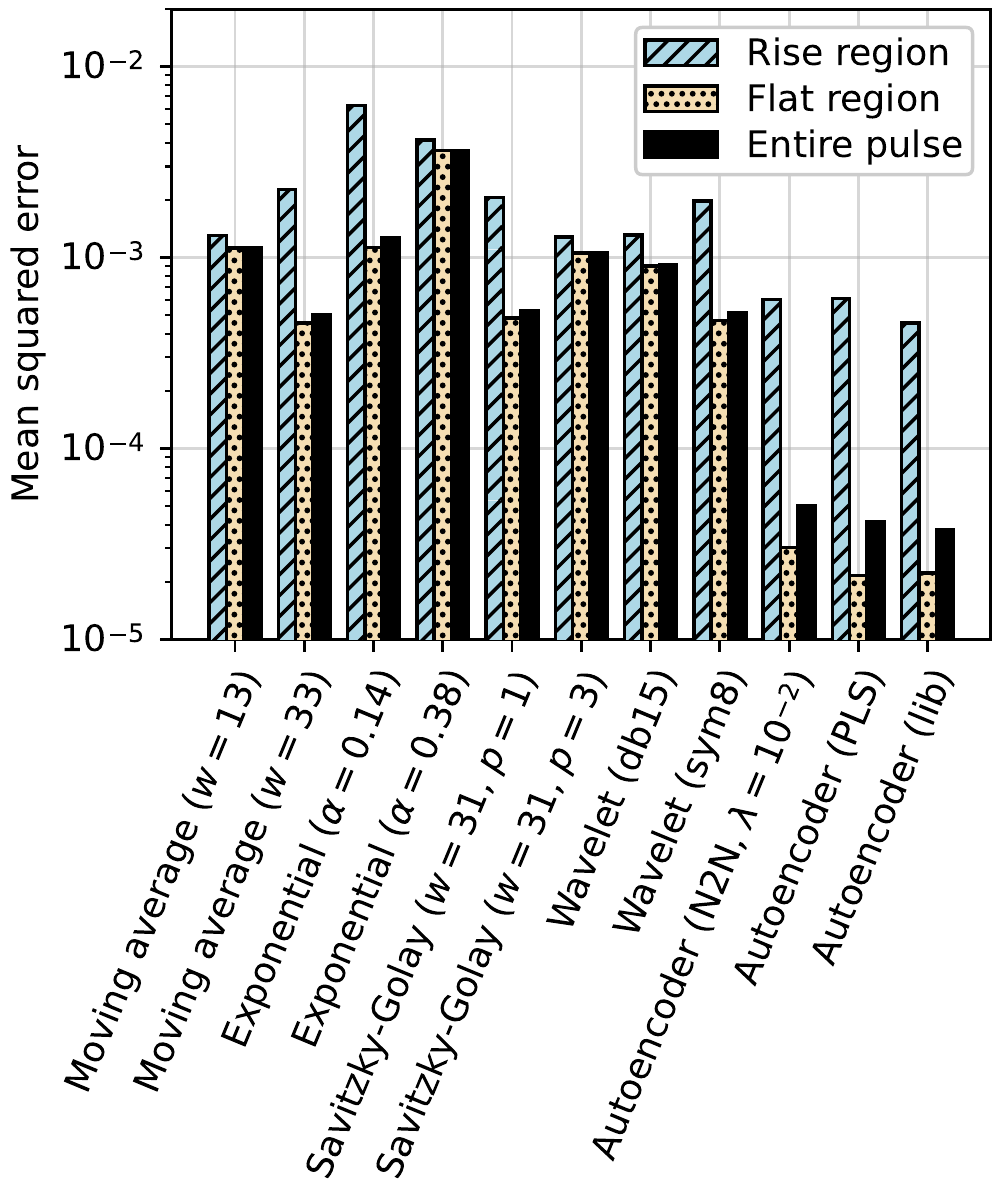}
    \caption{Mean squared error comparison of different noise removal methods inside and outside the rise region of the pulse (defined in the text and shown in \cref{fig:lib_pulses_examples_denoised}). The mean squared error inside the rise region is indicated by the slanted line hatch while outside the rise region is indicated by the dotted hatch. Solid fill corresponds to the mean squared error over the entire pulse.}
    \label{fig:mse_comparison}
\end{figure}

The mean squared error comparison between the traditional methods in \cref{sec:traditional_denoising_methods} is shown in \cref{fig:mse_comparison}. The performance of two autoencoder models trained with the regular procedure (one using the library set and one using the PLS set) and one model trained with the Noise2Noise procedure (using the $^{60}$Co data) are included. \Cref{fig:mse_comparison} contains two variations of each traditional method: one using the optimal parameter(s) for the rise region and one using the optimal parameter(s) for the entire pulse, both optimized on the validation set by minimizing the mean squared error. For wavelet-based methods, we used VisuShrink to determine a global threshold and soft thresholding to shrink the coefficients towards zero.

In addition to the four methods in \cref{fig:mse_comparison}, a Kalman filter was implemented using the denoised output from the autoencoder as the underlying model and the baseline RMS of the noisy pulse as the measurement error. The filter was optimized by selecting the level of extrapolation between the noisy data and the underlying model that resulted in the least error. This process resulted in an optimized model with zero extrapolation, representing a copy of the autoencoder model, and so the Kalman filter was omitted from \cref{fig:mse_comparison}.

The autoencoder outperformed all traditional methods in both the rise region and flat region of the pulses. While the method requires training to denoise a specific type of data, it does offer improvements over traditional denoising, as evaluated using the mean squared error. The structural similarity index measure (SSIM) \cite{wang2004image} was also used to compare the performance of the traditional denoising methods as it is designed to quantify image degradation with reference to human perception. However, the relative performance of each method using SSIM is nearly identical to that using the mean squared error, and so the corresponding SSIM comparison figure is not shown.

\subsubsection{Energy resolution comparison}
\label{sec:energy_resolution_comparison_sim}

In this section, we use simulations to calculate the energy resolution before and after denoising using the amplitude normalized library pulse model. For a given pulse, the energy is calculated from the amplitude of a trapezoidal filter with a given gap and shaping time. The energy resolution is then defined as the full width at half maximum (FWHM) of the energy peak.

\begin{figure}[ht]
    \centering
    
    \includegraphics[width=\properwidth]{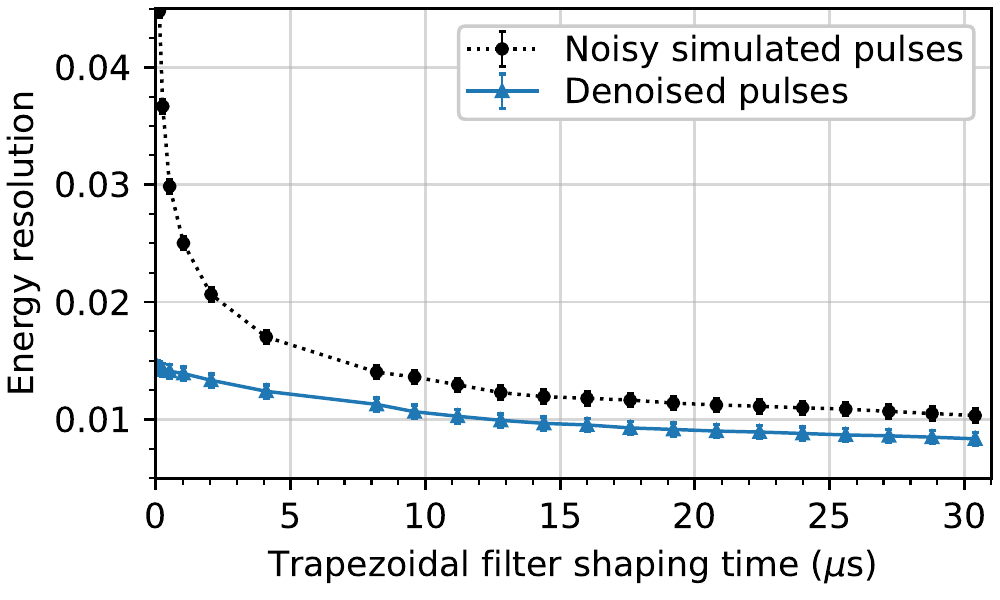}
    \caption{Energy resolution, defined as the FWHM of the energy distribution, as a function of trapezoidal filter shaping time. Calculated on library pulses with real detector noise with a baseline RMS of 0.1 before denoising (dotted line, circle markers) and after denoising with the amplitude normalized model (solid line, triangle markers).}
    \label{fig:lib_fwhm}
\end{figure}

We use the set of library pulses and real detector noise, distinct from the training set, to create new sets of \num{172400} noisy single-site event pulses for evaluation. A dataset is created for noise levels ranging from 0.02 to 0.2 in increments of 0.02. The energy resolution as a function of the trapezoidal filter shaping time on one such simulated dataset with a noise level of 0.1 is shown in \cref{fig:lib_fwhm} with and without denoising. This noise level is roughly the same as we observe with $^{241}$Am gamma rays. For all datasets, the gap time in the trapezoidal filter is fixed at $\SI{1.8}{\mu s}$, which was found to be sufficiently large for even the slowest rise times. Clean library pulses all have an amplitude of unity before noise addition, and so the ``true energy'' is one, meaning that the noise level and the resolution in \cref{fig:lib_fwhm} can both be interpreted as a fraction of the amplitude.

\begin{figure}[ht]
    \centering
    
    \includegraphics[width=\properwidth]{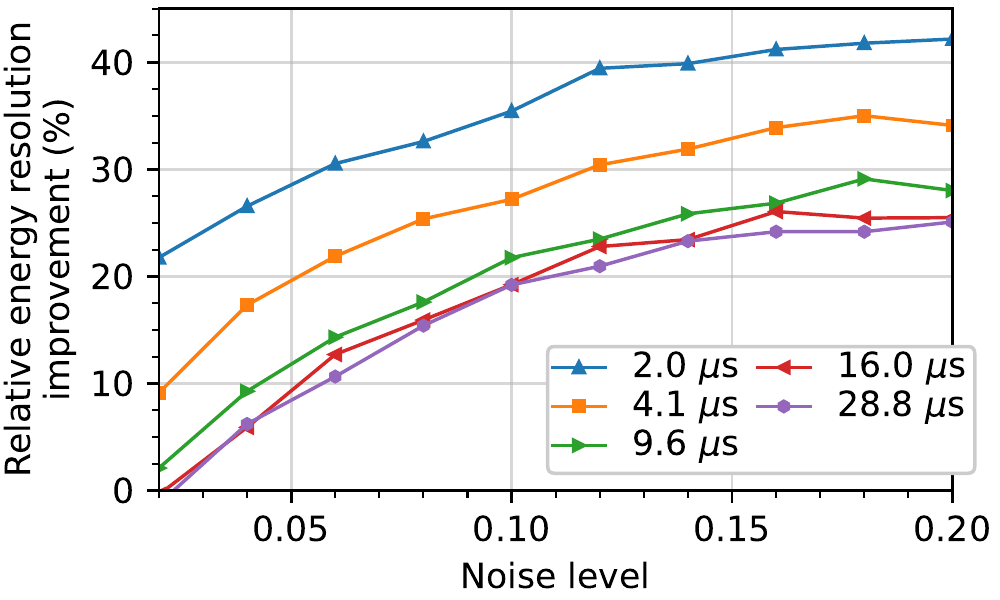}
    \caption{Relative improvement in the energy resolution from denoising as a function of noise level on simulated pulses. Each curve corresponds to a single trapezoidal filter shaping time.}
    \label{fig:lib_fwhm_rel_improvement}
\end{figure}

\Cref{fig:lib_fwhm_rel_improvement} shows the relative improvement in the energy resolution after denoising as a function of the noise level. Each curve corresponds to a given shaping time, and each point on a given curve corresponds to a point from a plot such as \cref{fig:lib_fwhm}.

At every shaping time, the energy resolution on simulated data is lower with denoising. This is particularly prominent at shaping times less than $\SI{5}{\mu s}$, where the improvement in energy resolution exceeds 20\% for all but the lowest noise levels. At higher shaping times, the averaging window of the trapezoidal filter is large enough to smooth out the noise, and so the improvement is generally smaller. Overall, the best improvement in energy resolution is obtained when the pulses have a high level of noise and when the shaping time used to calculate the energy is small. However, we observe improvements in the energy resolution resolution at every noise level and shaping time evaluated.

\subsection{Evaluation on detector data}

\begin{figure}[ht]
        \centering
        
        \includegraphics[width=\properwidth]{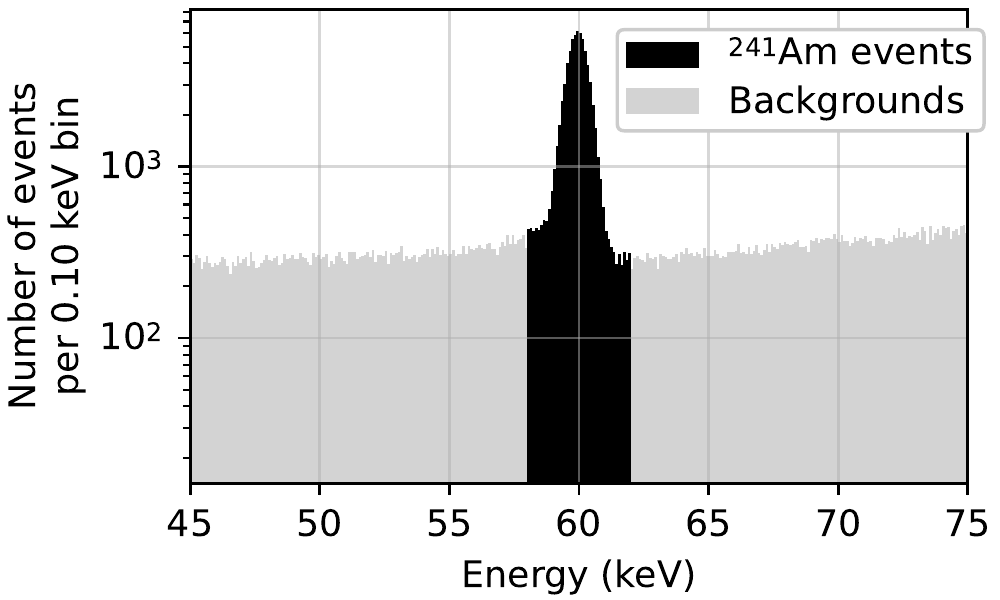}
        \caption{Amplitude distribution (converted to energy) of the $^{241}$Am data, as calculated from the trapezoidal filter. Events in the dark region are within $\pm\SI{2}{keV}$ of the peak and taken to be signal events from the $^{241}$Am source, while events in the light region are taken to be backgrounds. Note that the y-axis is logarithmic.}
        \label{fig:data_energy_distribution}
\end{figure}

In this section, we evaluate the denoising performance using data collected with the $^{241}$Am source, as it provides mono-energetic pulses with a reasonable amount of noise to evaluate the autoencoder. We select events within $\pm\SI{2}{keV}$ of the $\SI{60}{keV}$ peak associated with the source gamma rays, as seen in \cref{fig:data_energy_distribution}. The event energies are estimated using a trapezoidal filter. Pulses with outliers in the slopes on either the baseline or end of the trace are removed from the dataset.

Furthermore, the $\SI{60}{keV}$ events from the source are essentially all single site events for which the underlying true shape is close to the ones simulated in the library dataset. This allows us to infer a reasonable guess of the clean target by selecting the library pulse that minimizes the $\chi^2$ value defined in \cref{eqn:chi-square}. Note that the basis set used here was generated with a preamplifier time constant of $\SI{20}{ns}$ as that was found to best match our detector.

\subsubsection{Qualitative evaluation}

\begin{figure}[ht]
    \centering
    
    \includegraphics[width=\properwidth]{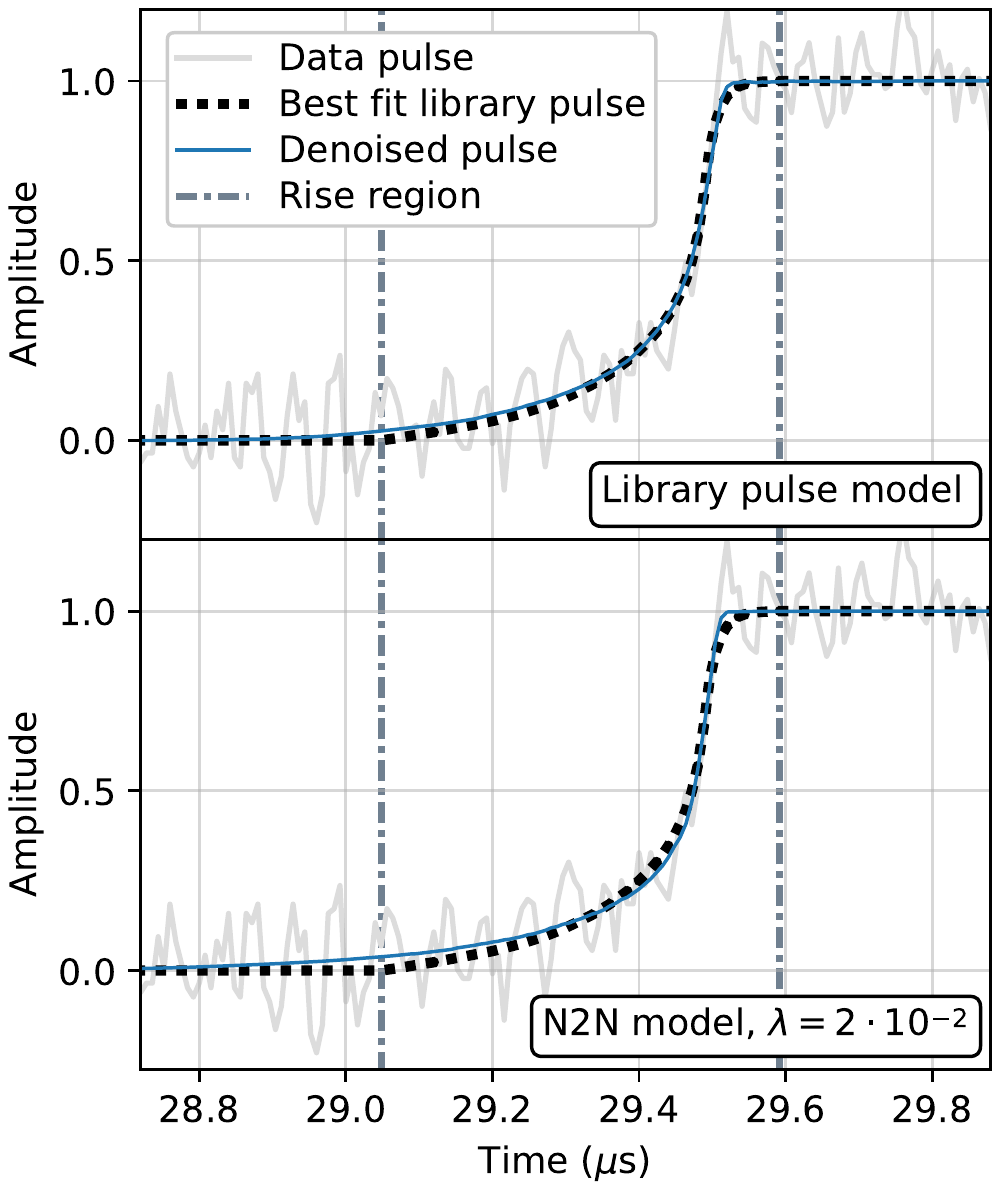}
    \caption{An example single-site event from the $^{241}$Am dataset. Included in each plot is the noisy data pulse (solid light line), the best fit library pulse (dotted line), and the corresponding denoised pulse (solid dark line) from the regular library pulse model (top) and Noise2Noise model (bottom).}
    \label{fig:data_pulses_sse_examples_denoised}
\end{figure}

When evaluating the performance on real data, one does not have a true clean signal with which to compare, as was the case with simulated pulses. Instead, we conduct several studies to understand how the autoencoder affects the shape of the denoised pulses. We start with a qualitative evaluation on a data pulse. \Cref{fig:data_pulses_sse_examples_denoised} shows a single-site event from the $^{241}$Am source data that has been denoised by two versions of our model. The top panel shows the pulse denoised with the regular library pulse model while the bottom panel shows the denoised pulse using the Noise2Noise model trained with $^{60}$Co data with a total variation penalty. Each plot also includes the library pulse with the lowest chi-squared value as a best estimate of the true underlying pulse.

Qualitatively, we observe that the performance of all models on the single-site event are promising and tend to better capture the shape of the pulse than the best fit library pulse. Furthermore, while not shown here, the denoising performance of all models on the multi-site events present in the set is also very good. As was observed with the results on simulated data, the library pulse model tends to perform the best. For the Noise2Noise models, the total variation penalty is important as the model trained without it retains some of the noise.

\subsubsection{Chi-squared comparison}

We compare the $\chi^2$ value between the data pulse and denoised pulse for events in the $^{241}$Am dataset. Although we do not know the true shape of the data pulses without noise, we can use the $\chi^2$ distribution to determine if, statistically, the denoised pulses are consistent with their noisy progenitors. It is important to use the $\chi^2$ instead of the mean or sum of squared errors to account for the signal-to-noise ratio of the pulses.

\begin{figure}[ht]
    \centering
    
    \includegraphics[width=\properwidth]{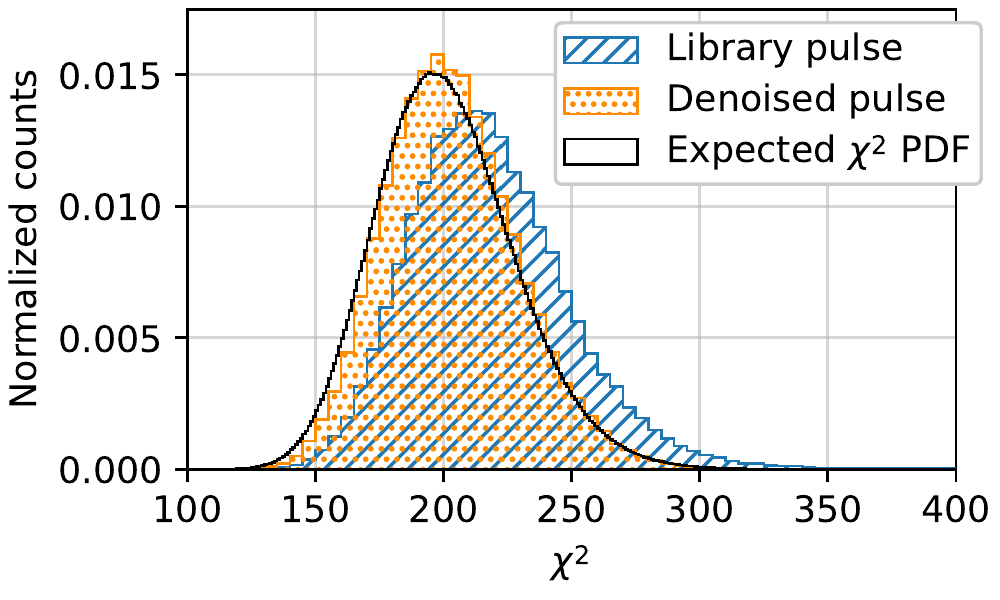}
    \caption{$\chi^2$ distribution computed between the data and both the denoised pulse (slanted line hatch) and best fit library pulse (dotted hatch).}
    \label{fig:data_chi2_selected}
\end{figure}

\Cref{fig:data_chi2_selected} shows the distribution of $\chi^2$ values between the data pulses and corresponding denoised pulses using the regular library noise removal model. It also includes the distribution of $\chi^2$ values between the data pulses and corresponding best matching library pulses as determined by the $\chi^2$-minimization across all library pulses in the basis set.

The $\chi^2$ value for each pulse is computed from sample 3600 to 3800, meaning that the number of degrees of freedom (NDF) is 200. This range is chosen because the pulses are horizontally centred within this window and it is long enough to capture the important components of the pulse for the largest rise times. A modified $\chi^2$ distribution for $\text{NDF}=200$ expected from our detector noise is also overlayed on \cref{fig:data_chi2_selected} for comparison. This distribution is calculated directly on the detector noise, independent of any signal. Note that since the detector noise is not Gaussian, the modified $\chi^2$ distribution is shorter and wider than the true $\chi^2$ probability density function for $\text{NDF}=200$ which assumes normally distributed noise.

The results show that the denoised pulses fit the data better than the best matching library pulses, as indicated by the lower mean and the better fit of the distribution to the $\chi^2$ distribution expected from the detector noise. This could indicate that there are not enough library pulses in the basis set, or that some other parameter of the detector is being modelled improperly. Additionally, while not shown here, the fit is better on \emph{all} events, not just those in the energy range within the $\SI{60}{keV}$ peak, indicating that our model is properly denoising multi-site events and low-energy events for which the pulse fitter fails to converge.

\subsubsection{Energy resolution comparison}

\begin{figure}[ht]
    \centering
    
    \includegraphics[width=\properwidth]{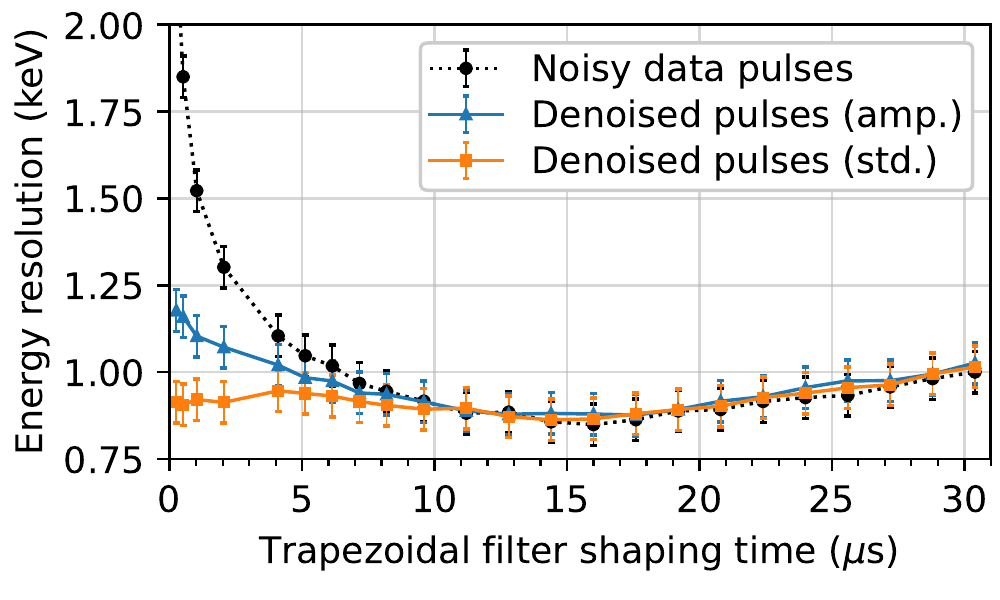}
    \caption{Energy resolution, defined as the FWHM of the energy distribution, as a function of trapezoidal filter shaping time. Calculated on data from the $^{241}$Am source before denoising (dotted line, circle markers) and after denoising with the amplitude normalized model (solid line, triangle markers) and standardized model (solid line, square markers).}
    \label{fig:data_fwhm}
\end{figure}

In this section, we compare the energy resolution of the $\SI{60}{keV}$ peak before and after denoising, similar to what was done in \cref{sec:energy_resolution_comparison_sim} on simulated data. The results are shown in \cref{fig:data_fwhm}, which illustrates the energy resolution on both the original noisy and denoised pulses as a function of the trapezoidal filter shaping time. The gap time in the trapezoidal filter is again fixed at $\SI{1.8}{\mu s}$. The results for two denoising models (both trained with the regular procedure and library pulses) are included: one using amplitude normalized pulses and the other using standardized pulses.

While denoising does not achieve a lower energy resolution overall, it does allow us to obtain a comparable energy resolution with a lower shaping time. This is especially true for the model trained with standardized pulses, where the optimal energy resolution is achieved even at shaping times under $\SI{2}{\mu s}$. The discrepancy between the two denoised models at lower shaping times is surprising as the only difference is the preprocessing of the pulses. However, all data pulses are horizontally centred, meaning that the lack of robustness to horizontal shifts for standardized pulse models is not an issue here. We note that in general, models trained with standardized pulses performed similarly in terms of both mean squared error (on simulations) and $\chi^2$ (on data), and the energy resolution is the only metric where a significant performance deviation is observed.

These results are important, as the optimal shaping time required to obtain a reasonable energy resolution directly affects data collection. Specifically, a longer shaping time requires longer traces, which in turn occupies more disk space and increases the chances of event pile-up. A lower shaping time thus has many practical implications with regards to easier/more efficient data storage and analysis. This is also of interest to higher rate applications of HPGe detectors, rather than just in rare-event searches.

As shown in \cref{sec:energy_resolution_comparison_sim}, we expect an overall improvement in the energy resolution from simulations under ideal conditions. However, we do not observe this on the real $^{241}$Am data. As well, the shape of the energy resolution curve is different; specifically, it decreases as a function of the shaping time up until the largest shaping time, which is limited by the length of the noise traces. This is in contrast to \cref{fig:data_fwhm}, where there is a clear minimum at around $\SI{15}{\mu s}$. A prominent imperfection with data is that there are multiple sources of exponential decay, while the pole zero correction only accounts for one “effective” source. This results in visually imperceptible changes to the ends of the pulses that broaden the energy resolution, particularly at high shaping times. As the models are not trained to remove this effect, it is still present to some extent even after denoising. This was confirmed by adding multiple exponential decays and then applying only one correction to a simulated test set. However, as we do not know the properties of the exponential decays present in our setup, we have not accounted for this in the training data or removed it from the real data.

\subsubsection{Frequency spectra comparison}

The frequency power spectrum of denoised pulses was analyzed for a subrun of the $^{241}$Am dataset described in \cref{sec:experiment_detector_data} where the source was fixed at roughly half the height of the detector. The data were denoised with some of the traditional denoising methods evaluated in \cref{sec:denoising_methods_comparison}, as well as with some of our autoencoder models. The discrete Fourier transform was computed on these denoised data in addition to the original noisy pulses in the subrun. In computing the discrete Fourier transform, a Hanning window \cite{blackman1958measurement} was used in order to remove discontinuities resulting from the step-like shape of the pulses and to lessen spectral leakage.

\begin{figure}[ht]
    \centering
    
    \includegraphics[width=\properwidth]{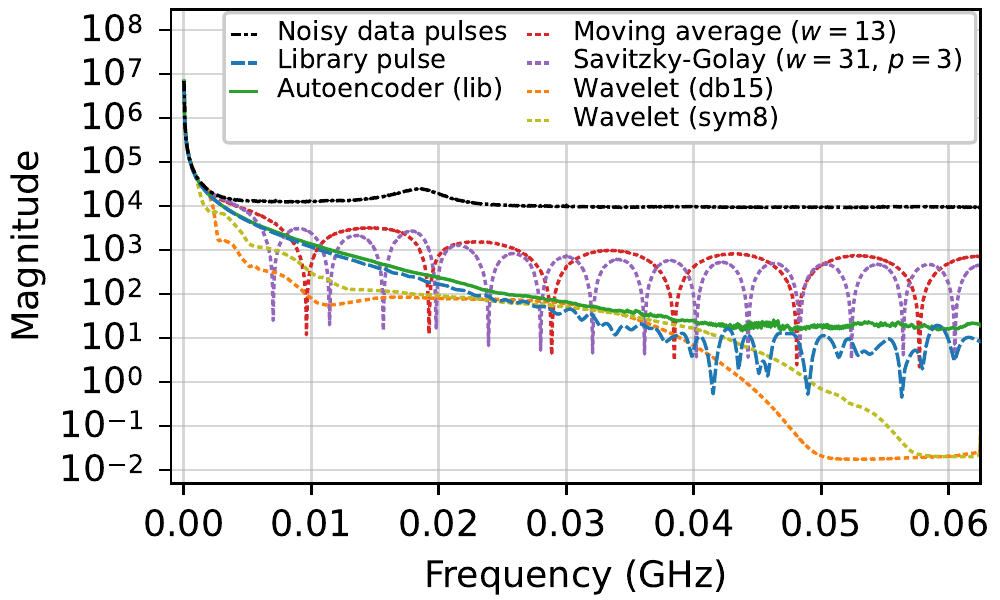}
    \caption{Comparison of the frequency spectra for simulated library pulses, noisy pulses, pulses denoised with the autoencoder trained using the regular procedure and library pulses, and pulses denoised with four different traditional denoising methods.}
    \label{fig:fourier_traditional}
\end{figure}

\begin{figure}[ht]
    \centering
    
    \includegraphics[width=\properwidth]{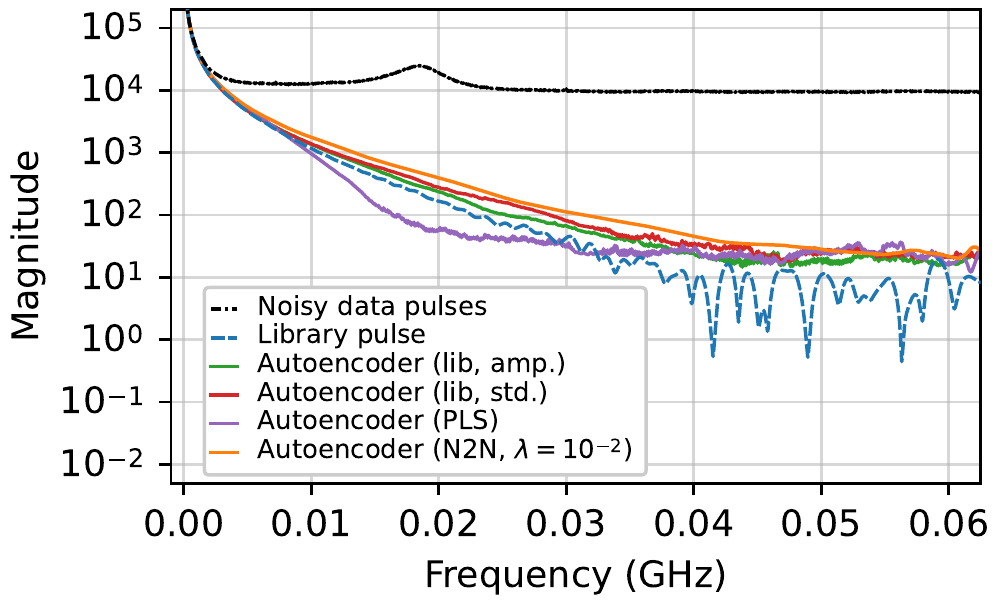}
    \caption{Comparison of the frequency spectra for simulated library pulses, noisy pulses, and pulses denoised with four different autoencoder models.}
    \label{fig:fourier_autoencoder}
\end{figure}

The resulting frequency spectra, obtained by summing the individual spectrum of each pulse for a given set of data, are shown in \cref{fig:fourier_traditional} for the moving average, Savitzky-Golay, and wavelet denoising methods (all using the optimized parameters as in \cref{fig:mse_comparison}). \Cref{fig:fourier_autoencoder} shows the frequency spectra comparisons for regular autoencoder models trained with library and PLS simulated data and the Noise2Noise model trained with real data. It also includes the regular library pulse autoencoder trained with standardized pulses. In each figure, the frequency spectra for the noisy data and a clean simulated pulse at the position in the detector corresponding to the location of the $^{241}$Am source are also shown for comparison.

Of the methods evaluated, the pulses denoised by the autoencoders have the closest frequency spectra to that of the clean simulated pulse. In \cref{fig:fourier_traditional}, both the moving average and Savitzky-Golay method spectra have periodic artifacts from their windows. Both wavelet denoising method spectra tend to be overly aggressive in removing noise, specifically at lower frequencies (up to approximately $\SI{0.02}{GHz}$) and the highest frequencies (greater than $\SI{0.05}{GHz}$).

In \cref{fig:fourier_autoencoder}, both regular autoencoder models trained with library pulses and the Noise2Noise model have similar frequency spectra. The Noise2Noise model has the smoothest spectrum, but is less aggressive in removing noise at virtually all frequencies, particularly at frequencies larger than $\SI{0.04}{GHz}$ and around $\SI{0.06}{GHz}$, indicating that it is fails to remove some of the higher frequency noise. This is consistent with visual observations of the denoised pulses. The autoencoder trained with PLS pulses has the most divergent spectrum and is removing some portion of the signal at the mid-range of frequencies. It does, however, perform similarly to the library pulse models at higher frequencies. This could be due to the fact that the PLS pulse model is more likely to distort and ``smooth out'' the pulse in and around the rise region, while in principle it should denoise the flat regions just as well as the library pulse models. These results are in agreement with those from \cref{fig:mse_comparison}, which shows that while the Noise2Noise model is comparable or slightly better at preserving features in the rise region, it is inferior at denoising the overall pulse.

\section{Conclusions}
\label{sec:conclusion}

We have applied deep convolutional autoencoders to strongly suppress electronic noise from one-dimensional signals collected with a p-type point contact high purity germanium detector, while demonstrating that the underlying pulse shape is preserved to a high degree of accuracy. Comparisons with simulated data, which allow for the underlying clean signal to be known, show that the autoencoder outperforms various traditional denoising methods in this task. We also showed the excellent performance of the autoencoder using real detector data. We found that the denoised pulses are statistically consistent with the original noisy pulses collected from an $^{241}$Am calibration source. Notably, the distribution of chi-squared values made from the denoised pulses is in better agreement with the expected distribution than that made with the best fitting simulated pulses obtained from detailed physics simulations. We also showed that denoising allows one to reach the optimal energy resolution with a trapezoidal filter at significantly reduced shaping times for low energy pulses. This has practical implications on data collection, storage, and analysis as it can allow for shorter traces to be collected without loss in energy resolution. We expect that, under certain circumstances, the energy resolution at low energies can even be reduced with denoising, in particular if any residual exponential decays in the pulse shape can be removed.

In addition to the excellent performance of the models trained on library pulses, we have presented two methods to train models that can perform well without the need for detailed detector simulations. We created a set of pulses using piecewise linear functions with smoothing that are qualitatively similar to the library pulses. Models trained with these pulses performed almost as well as models trained with the library dataset. Additionally, we used a modification to an existing procedure – the Noise2Noise method – to train an autoencoder with no underlying clean pulse as a basis of truth. Instead, the model only required data and noise collected from the detector. While it did not outperform any of the standard models trained with simulated pulses, it still outperformed traditional denoising techniques such as wavelet-based methods. The Noise2Noise method is most limited by the amount and quality of data used to train it. More data, particularly from different sources over a wider energy range, would allow the network to see more variation in pulses and presumably improve the model.

Although we demonstrate the performance of the convolutional autoencoder on germanium detector data, there are many applications that can benefit from the removal of electronic noise in one-dimensional signals. We are beginning to apply our methods to other detector technologies such as gaseous proportional counters and bubble chambers with promising results. Furthermore, denoising autoencoders can be used to obtain a more robust latent representation of the original input \cite{vincent2010stacked}. We are exploring the use of the encoder portion of the network for other tasks, including clipping restoration, peak finding, and single-site/multi-site event discrimination (similar to the work in \cite{holl2019deep}). Our work has great potential to be expanded upon, and the methods demonstrated here are broadly applicable to not only the particle astrophysics community, but any field dealing with noisy one-dimensional signals.

%%===============================================%%
%% Acknowledgements, statements and declarations %%
%%===============================================%%

\section*{Acknowledgments}

This work was primarily supported by the Natural Sciences and Engineering Research Council of Canada, funding reference numbers SAPIN-2017-00023 and CGSD3-546735-2020. We acknowledge the support of the Arthur B. McDonald Canadian Astroparticle Physics Research Institute through the Highly Qualified Personnel pooled funding opportunity, the Canada Foundation for Innovation John R.~Evans Leaders Fund, and the Walter C. Sumner Memorial Fellowship. We also thank the NVIDIA corporation for their support through their academic hardware grant program.

\section*{Statements and declarations}

\begin{itemize}
    \item \textbf{Competing interests}: The authors have no competing interests to declare that are relevant to the content of this article.
    \item \textbf{Data availability}: The datasets generated and analysed during the current study are available from the corresponding author on reasonable request.
    \item \textbf{Publication}: This version of the article has been accepted for publication, after peer review but is not the Version of Record and does not reflect post-acceptance improvements, or any corrections. The Version of Record is available online at:

    \href{https://doi.org/10.1140/epjc/s10052-022-11000-w}{\nolinkurl{doi.org/10.1140/epjc/s10052-022-11000-w}}
\end{itemize}

%%==============%%
%% Bibliography %%
%%==============%%

\bibliography{denoise}

\end{document}